\documentclass[aps,twocolumn,floats,prd,nofootinbib,superscriptaddress]
{revtex4-2}

\usepackage{amsmath,amsfonts,amssymb}
\usepackage{graphicx}
\usepackage{mathrsfs}
\usepackage{xcolor}
\usepackage{bm}
\usepackage{hyperref}
\usepackage{geometry}
\usepackage{booktabs}

\hypersetup{
    colorlinks=true,
    linkcolor=red,
    filecolor=magenta,
    urlcolor=blue,
    citecolor=blue,
}

\newcommand{\Mh}{M_h}
\newcommand{\SFR}{\dot{M}_\star}
\newcommand{\meanSFR}{\overline{\dot{M}}_\star}
\newcommand{\dn}{\frac{dn}{d\Mh}}

\newcommand{\Myr}{\,{\rm Myr}}
\newcommand{\Msun}{M_\odot}

\newcommand{\sigxiline}{\sigma_{\xi,\lambda}}
\newcommand{\Bperhalo}{B_\lambda}
\newcommand{\sigPS}{\sigma_{\rm PS}}
\newcommand{\tauPS}{\tau_{\rm PS}}
\newcommand{\Pshot}{P_{\rm shot}}
\newcommand{\Lbar}{\bar{L}}
\newcommand{\Ibar}{\bar{I}}
\newcommand{\beff}{b_{\rm eff}}
\newcommand{\Pm}{P_m}
\newcommand{\Halpha}{{\rm H}\alpha}
\newcommand{\CII}{[{\rm C}\,\textsc{ii}]}
\newcommand{\Lya}{{\rm Ly}\alpha}
\newcommand{\be}{\begin{equation}}
\newcommand{\ee}{\end{equation}}
\begin{document}

\title{When galaxies burst: enhanced shot-noise for line-intensity mapping in the JWST era}

\author{Ely D. Kovetz}
\email{kovetz@bgu.ac.il}
\affiliation{Department of Physics, Ben-Gurion University of the Negev, Be’er Sheva 84105, Israel}

\author{Hovav Lazare}
\affiliation{Department of Physics, Ben-Gurion University of the Negev, Be’er Sheva 84105, Israel}

\author{Sarah Libanore}
\affiliation{Department of Physics, Ben-Gurion University of the Negev, Be’er Sheva 84105, Israel}

\author{Julian B. Mu\~noz}
\affiliation{Department of Astronomy, The University of Texas at Austin, Austin, TX 78712, USA}
\affiliation{Cosmic Frontier Center, The University of Texas at Austin, Austin, TX 78712, USA}

\author{Eleonora Vanzan}
\affiliation{School of Physics and Astronomy, Tel-Aviv University, Tel-Aviv 69978, Israel}
\affiliation{Department of Physics, Ben-Gurion University of the Negev, Be’er Sheva 84105, Israel}

\begin{abstract}
Recent JWST observations indicate that star formation at $z\!\sim\!4-6$ is  more stochastic than previously assumed, with rms log-SFR scatter $\sim\!0.6$\,dex at $\Mh\!\sim\!10^{11}\Msun$, growing toward smaller halos and time-correlated on $\sim\!25$\,Myr.  This is significantly  higher than the typical $\sim\!0.3$\,dex phenomenological lognormal scatter assumed in standard line-intensity mapping (LIM) forecasts. 
We propagate the JWST-era burstiness through to the LIM shot-noise power spectrum and show that the result is a simple multiplicative correction: the deterministic shot noise multiplied by a line-dependent boost factor $\mathcal{B}_\lambda$ derived in closed form by convolving the SFR correlation function with the stellar-population-synthesis kernel of each line. At $z\!\sim\!6$, we find $\mathcal{B}_{\Halpha}\!\simeq\!7$ and $\mathcal{B}\!\sim\!2.5$--$3.5$ for longer-window tracers ($\CII$, CO, UV) --- factors of $\sim\!2$--$5$ above the standard prescription, and growing further toward higher redshift. The enhancement transforms the LIM landscape: it improves auto-spectrum detectability and suppresses lower-redshift interloper contamination, but degrades cosmological applications such as BAO that rely on a clean clustering measurement. Crucially, it also opens a new use of LIM as a diagnostic of high-redshift star-formation physics beyond the regime of individually resolved galaxies: \emph{redshift tomography} of a single line constrains the amplitude and mass dependence of the burstiness, while \emph{cross-line shot-noise correlations} probe its time coherence.
\end{abstract}

\maketitle

\noindent\textit{Introduction.}~%
\emph{JWST} is rapidly transforming our picture of galaxy formation
in the first billion years.  The high-redshift universe it has
revealed is more complicated than expected: the UV luminosity
function at $z\!\gtrsim\!10$ is brighter than pre-launch models
predicted~\cite{Mason:2023,Harikane:2023,Finkelstein:2024,Donnan:2024,Robertson:2024}, 
the $\Halpha$/UV ratio shows a wide
spread at $z\!\sim\!4$--$6$~\cite{Endsley:2024,Munoz:2026}, and individual
galaxies appear to undergo dramatic short-timescale fluctuations in
their star-formation rates~\cite{Sun:2023b,Gelli:2024,Pallottini:2023,Mitsuhashi:2025,Clarke:2024}.  
A common thread is the notion of \emph{burstiness}: that star formation
in early galaxies, especially in low-mass halos where stellar
feedback is most disruptive, proceeds in episodic bursts rather than
smoothly.  This has long been a generic prediction of
feedback-regulated star formation, and recent work has begun to turn
it from a qualitative idea into a quantitative, empirically anchored
description~\cite{Hopkins:2014,Sparre:2017,ElBadry:2016,FaucherGiguere:2018,Tacchella:2020,FurlanettoMirocha:2022,Munoz:2023clustering,Sun:2024framework,Munoz:2026}.

Bursty star formation has profound implications for line-intensity
mapping (LIM), the technique of measuring the aggregate redshifted
line emission from all sources in cosmic
volumes~\cite{Bernal:2022,Kovetz:2017}.  The observable
LIM power spectrum decomposes into clustering and shot-noise
terms,
\be
P_\lambda(k) = \Ibar_\lambda^2\,\beff^2(\lambda)\,\Pm(k) + \Pshot^\lambda\,,
\label{eq:Pk_decomp}
\ee
where $\Ibar_\lambda$ is the mean intensity in band $\lambda$,
$\beff(\lambda)$ the luminosity-weighted halo bias, $\Pm(k)$ the
linear matter power spectrum, and $\Pshot^\lambda$ integrates 
the second moment of $L_\lambda(M_h)$, the 
luminosity--halo-mass relation~\cite{Schaan:2021}, 
\be
\Pshot^\lambda = \mathcal{X}_\lambda^2\!\int d\Mh\,\dn\,\langle L_\lambda^2\rangle(\Mh) = \Pshot^\lambda\big|_{\rm det.} + \Pshot^\lambda\big|_{\rm stoch.}\,,
\label{eq:Pshot_decomp}
\ee
where $dn/dM_h$ is the halo mass function (HMF) and 
$\mathcal{X}_\lambda\!\equiv\!cy_\nu/(4\pi)$ is the standard 
luminosity ($L_\lambda$) to intensity ($I_\lambda$) conversion factor 
($c$ is the speed of light and $y_\nu\!\equiv\!d\chi/d\nu_{\rm obs}$ the 
comoving distance per unit observed frequency). The decomposition 
$\langle L_\lambda^2\rangle\!=\!\Lbar_\lambda^2 + \mathrm{Var}(L_\lambda)$ 
at fixed $\Mh$ separates two physically distinct contributions: 
the halo-discreteness term~\cite{Schaan:2021}
\be
\Pshot^\lambda\big|_{\rm det.} = \mathcal{X}_\lambda^2\!\int d\Mh\,\dn\,\Lbar_\lambda^2(\Mh)\,,
\label{eq:Pshot_det}
\ee
which arises purely from halo counting Poisson statistics under a 
deterministic luminosity--halo-mass relation, and the galaxy-stochasticity 
term $\Pshot^\lambda|_{\rm stoch.}\!=\!\mathcal{X}_\lambda^2\!\int d\Mh\,\dn\,\mathrm{Var}(L_\lambda)(\Mh)$, 
which captures fluctuations in the per-halo luminosity at fixed $\Mh$. 
Standard LIM forecasts retain only the halo-discreteness 
term~\cite{Lidz:2011,Pullen:2013,Breysse:2014,Breysse:2015saa,Yue:2015}, 
occasionally augmenting it with a phenomenological log-normal 
 scatter $\sigma_L\!\sim\!0.3$--$0.5$~dex at fixed $\Mh$ 
calibrated to simulations~\cite{Sun:2019,Yang:2022};  this $\sigma_L$ captures the combined 
effect of multiple sources of scatter (feedback variability, halo assembly history, AGN~\cite{Iyer:2020}), but 
without direct calibration against burstiness-sensitive observables at high reshift.

The recent analysis by Mu\~noz et al.~\cite{Munoz:2026} (M26) changes
this.  By combining the UV luminosity function, galaxy clustering,
and $\Halpha$/UV ratio data, M26 infer a burstiness amplitude (the
square root of the integrated star formation rate power spectrum, hence ``PS'')
$\sigPS\!\simeq\!2.0\!\pm\!0.3$ at a pivot mass $\Mh\!=\!10^{11}\Msun$\footnote{M26 quote $\sigPS$ at pivot $10^{10}\Msun$, where $\sigPS\!\simeq\!2.5\!\pm\!0.5$ depending on dataset (this also allows to fit high-z UVLFs~\cite{Munoz:2026, Sun:2023b,Gelli:2024}); propagating via their mass-dependence slope  to $10^{11}\Msun$ gives $\sigPS\!\simeq\!2.0\!\pm\!0.3$.} with
mass dependence $d\sigPS/d\log_{10}\Mh\!\simeq\!-0.5$, and a burst coherence
time  $\tauPS\!\simeq\!25^{+30}_{-10}$~Myr.  This corresponds
to a large rms log-SFR scatter $\sigma_x\!=\!\sigPS/\sqrt{2}\!\approx\!0.6$~dex
at the pivot mass, which  grows toward smaller halos,  precisely the
regime LIM is designed to probe.  
While these inferences are
model-dependent and should be tested against independent data,
the central values are large enough to warrant a careful revisit of LIM forecasts to account for burstiness.

In this work we propagate the M26 framework through to LIM observables. The
underlying idea is simple: if galaxies are bursty, their luminosities
at fixed halo mass are scattered around the mean, and
$\Pshot^\lambda$ no longer reduces to a $\Lbar_\lambda^2$ integral.
The size of the correction is set by the convolution of the SFR
correlation function with the time response of each line.  We carry
out this calculation analytically in the M26 effective burstiness
model, arriving at closed-form semi-analytic expressions for the auto and cross
LIM shot noise that factorize into the standard deterministic integral
times a line- and mass-dependent boost, and 
map out how the boost depends on  $\sigPS$ and $\tauPS$.  These expressions are
ready-to-use inputs for existing LIM forecasting pipelines~\cite{MasRibas:2023,Sun:2023LIMFAST,Munoz:2023Zeus,Libanore:2025},
subject to the line-modeling assumptions discussed below.
As we will demonstrate, the burstiness-induced enhancement of the shot noise 
improves detection prospects for the auto-power spectrum but degrades 
cosmological applications of high-redshift LIM that rely on a clean clustering measurement.

\smallskip
\noindent\textit{Per-halo boost.}~%
Adapting the M26 framework (see also~\cite{Caplar:2019,Sun:2024framework}, we model the SFR of a galaxy in a halo of mass $\Mh$ at cosmic time $t$ as a mean-anchored lognormal,
\be
\SFR(\Mh,t) = \meanSFR(\Mh,t)\,e^{x(\Mh,t)-\sigma_x^2(\Mh)/2}\,,
\label{eq:SFR_lognormal}
\ee
where $x$ is a zero-mean Gaussian field with stationary Ornstein--Uhlenbeck (OU) autocorrelation~\cite{UhlenbeckOrnstein:1930} $\xi_x(\Delta t)\!=\!\sigma_x^2 e^{-|\Delta t|/\tauPS}$ and $\sigma_x^2\!=\!\sigPS^2/2$.  To account for SFR history, the line luminosity is weighted by a stellar population synthesis (SPS)  Green's function~\cite{BruzualCharlot:2003,Munoz:2026}, $L_\lambda(\Mh)\!=\!\int dt_a\,G_\lambda(t_a)\SFR(\Mh,t_{\rm obs}\!-\!t_a)$. The shot noise, Eq.\eqref{eq:Pshot_decomp}, depends on $\langle L_\lambda^2\rangle\!=\!\Lbar_\lambda^2[1\!+\!V_\lambda]$, defining $V_\lambda\!\equiv\!{\rm Var}(L_\lambda)/\Lbar_\lambda^2$ as the dimensionless luminosity variance at fixed $\Mh$.

For a top-hat SPS window of effective width $t_\lambda$ (a useful proxy for the time over which each line responds to recent star formation~\cite{Munoz:2026,Sun:2019}), the dimensionless luminosity variance reduces to a single integral~(Appendix~B):
\be
V_\lambda(\Mh) = \frac{2}{t_\lambda^2}\!\!\int_0^{t_\lambda}\!\!\!ds\,(t_\lambda\!-\!s)\Big[e^{\sigma_x^2(\Mh) e^{-s/\tauPS}}\!-\!1\Big].
\label{eq:Vexact}
\ee
This expression is a closed-form, exact resummation of the lognormal-mapped Gaussian correlator $\xi_y(\Delta t)\!=\!\exp[\xi_x(\Delta t)]\!-\!1$ convolved with the line window, valid for any $\sigma_x^2$.  Two limits are useful: for $\tauPS\!\gg\!t_\lambda$ (long bursts, narrow window), $V_\lambda\!\to\!e^{\sigma_x^2}\!-\!1$; for $\tauPS\!\ll\!t_\lambda$ (broad window), $V_\lambda\!\simeq\!(2\tauPS/t_\lambda)\!\int_0^1\!du(e^{\sigma_x^2 u}\!-\!1)/u$, parametrically much larger than the linearized $(\tauPS/t_\lambda)\sigma_x^2$ scaling at large $\sigma_x$. Eq.~\eqref{eq:Vexact} agrees with direct Monte Carlo realizations of the OU process at the percent level (Appendix~B).

\smallskip
\noindent\textit{Boosted shot noise.}~%
  We define the per-halo boost $\Bperhalo(\Mh)\!\equiv\!1\!+\!V_\lambda(\Mh)\!=\!e^{\sigxiline^2(\Mh)}$ with $\sigxiline^2\!\equiv\!\ln[1\!+\!V_\lambda]$ playing the role of an effective lognormal scatter for the line (to be compared with the phenomenological $\sigma^2_L$ of standard LIM forecasts). The total shot noise, Eq.\eqref{eq:Pshot_decomp}, then becomes 
\be
\Pshot^\lambda = \mathcal{X}_\lambda^2 \int d\Mh\,\dn\,\Lbar_\lambda^2(\Mh)\,e^{\sigxiline^2(\Mh)}\,.\;
\label{eq:Pshot_main}
\ee
This is the central result of the paper.  Compared with the
deterministic (halo-discreteness) expression Eq.~\eqref{eq:Pshot_det}, the only change is
the per-halo enhancement factor $e^{\sigxiline^2(\Mh)}$ inside the
integrand.  Defining the mass-integrated boost,
\be
\mathcal{B}_\lambda \equiv \frac{\Pshot^\lambda}{\Pshot^\lambda|_{\rm det.}}
= \frac{\int d\Mh\,(dn/d\Mh)\,\Lbar_\lambda^2\,e^{\sigxiline^2(\Mh)}}
       {\int d\Mh\,(dn/d\Mh)\,\Lbar_\lambda^2}\,,
\label{eq:Blam}
\ee
$\mathcal{B}_\lambda$ is simply a $\Lbar_\lambda^2$-weighted average of
the per-halo enhancement.  Because $\sigPS$ grows toward low $\Mh$, the
shot-noise boost is weighted toward low-mass halos. This  makes LIM surveys uniquely sensitive
to the burstiness of faint, individually unresolvable galaxies at
high redshift.

Before we continue, it is worth emphasizing what this calculation does \emph{not} change.
The mean intensity
$\Ibar_\lambda\!=\!\mathcal{X}_\lambda\int (dn/d\Mh)\Lbar_\lambda$ and
the luminosity-weighted bias
$\beff(\lambda)\!=\![\int(dn/d\Mh)b_h\Lbar_\lambda]/[\int(dn/d\Mh)\Lbar_\lambda]$
in Eq.~\eqref{eq:Pk_decomp} are unchanged by burstiness in our
convention.  Note that the Green function $G_\lambda(t_a)$ does enter the clustering
term as well, through $\Lbar_\lambda(\Mh)\!=\!\int dt_a G_\lambda(t_a)\,\langle\dot M_\star\rangle(\Mh,t_{\rm obs}\!-\!t_a)$;
but because the mean-anchored convention Eq.~\eqref{eq:SFR_lognormal} sets $\langle\dot M_\star\rangle$ exactly equal
to its UVLF-calibrated value by construction, the OU process drops out of the first moment entirely.\footnote{We adopt the time-stationary approximation $\langle\dot M_\star\rangle(t_{\rm obs}-t_a)\!\approx\!\langle\dot M_\star\rangle(t_{\rm obs})$ over the support of $G_\lambda$, valid to a few percent at $z\!\sim\!6$ even for the longest UV window; under this approximation the Green function reduces to a line-dependent normalization in $\Lbar_\lambda$. See Appendix~B for the explicit calculation and further discussion.}
Burstiness only survives in the second moment $\langle L_\lambda^2\rangle$, where the OU correlator
sets $V_\lambda(\Mh)$. The asymmetry between clustering and shot noise is thus a feature of the
mean-anchoring choice, not an inconsistency: in the M26 \emph{median} convention $\Ibar_\lambda$ and $\beff$
would pick up factors of $e^{\sigma_x^2/2}$ that get absorbed into the star-formation efficiency (SFE) parameters, with identical
observable predictions.  The boost factor $\mathcal{B}_\lambda$ itself, though, is convention-independent. The LIM shot-noise enhancement we describe is a physical effect, not a parametrization choice.

\smallskip
\noindent\textit{Line-modeling.}~%
Throughout the body of this paper we adopt the minimal ansatz
$\Lbar_\lambda(\Mh)\!\propto\!\meanSFR(\Mh)$ with a line-independent
proportionality (per-line $\kappa_\lambda$ amplitudes only enter the
prefactor of $\Pshot^\lambda$ and $\Ibar_\lambda$, not  $\mathcal{B}_\lambda$).
This isolates the effect we wish to characterize, the propagation of
SFR burstiness through the SPS Green's function, with the line entering
only through its effective window width $t_\lambda$. Real LIM line
models  involve metallicity-, ISM-, and feedback-dependent shape
modifications of $\Lbar_\lambda(\Mh)$, and  some are  calibrated against
data that already contain  scatter, possibly leading to double-counting when our
boost is applied verbatim. We discuss line-modeling considerations,
and quantify the resulting shifts in $\mathcal{B}_\lambda$, in
Appendix A.

\smallskip
\noindent\textit{Numerical estimates.}~%
We evaluate $V_\lambda(\Mh)$ from Eq.~\eqref{eq:Vexact} for four representative LIM tracers using
top-hat proxies of effective width $t_\lambda$:
$\Halpha$ ($t_\lambda\!=\!7$~Myr, equivalent to BC03~\cite{BruzualCharlot:2003} e-folding time $t_e=t_\lambda/2\approx3.5$~Myr; see Appendix B),
UV continuum ($t_\lambda\!=\!100$~Myr, $t_e\approx50$~Myr), $\CII$ 158\,$\mu$m
($t_\lambda\!=\!50$~Myr~\cite{Lagache:2018}), and CO(1--0)
($t_\lambda\!=\!80$~Myr~\cite{Li:2016,Sun:2019}).  For
$\Halpha$ and UV the top-hat agrees with the realistic exponential Green's function
to $\lesssim\!7\%$ in $\sigxiline^2$ across the relevant mass range
(see Appendix~B for the matching prescription); for [\textsc{C\,ii}] and CO, where the line response depends on
ISM physics not captured by a single timescale, our quoted boosts are
indicative only.

\begin{figure}[t]
\centering
\includegraphics[width=0.925\columnwidth]{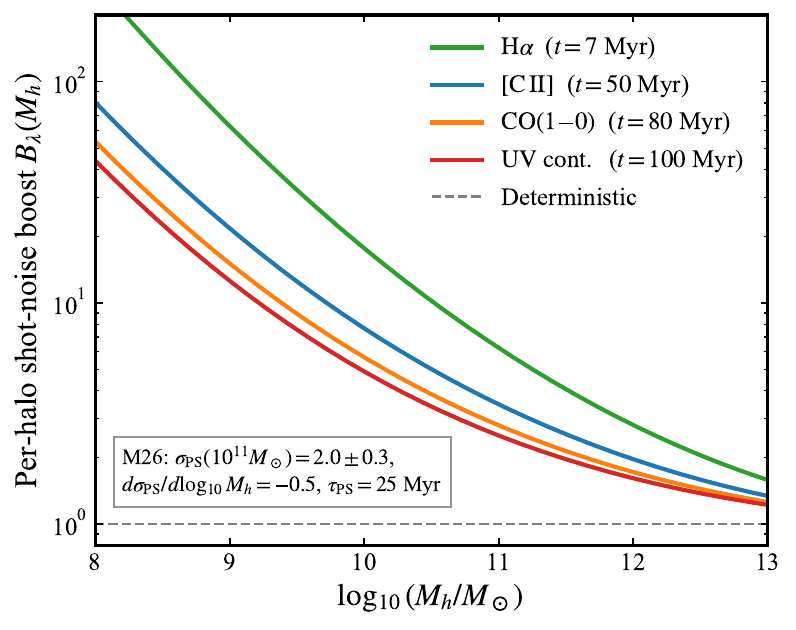}
\caption{\textbf{Per-halo shot-noise enhancement vs.\ halo
mass for four representative LIM tracers at $z\!=\!6$,}
$\langle L_\lambda^2\rangle/\Lbar_\lambda^2\!=\!e^{\sigxiline^2(\Mh)}$,
relative to the deterministic prediction (dashed line at unity), for
the M26 best-fit parameters
[$\sigPS(10^{11}\Msun)\!=\!2.0\!\pm\!0.3$,
$d\sigPS/d\log_{10}\Mh\!=\!-0.5$, $\tauPS\!=\!25$~Myr].  Lines
tracing recent star formation ($\Halpha$) experience the largest
per-halo enhancement, especially in low-mass halos where $\sigPS$ peaks.
}
\vspace{-0.15in}
\label{fig:boost}
\end{figure}

Fig.~\ref{fig:boost} shows the per-halo boost $e^{\sigxiline^2(\Mh)}$.  The mass
dependence is steep: $\Halpha$ shot noise per halo is enhanced by $\sim\!60$ at
$\Mh\!=\!10^9\Msun$, dropping to $\sim\!6$ at $\Mh\!=\!10^{11}\Msun$.  Even the
longer-timescale tracers experience per-halo enhancements of $\sim\!4$--$25$ at the low-mass
end (Tab.~\ref{tab:boost}).

\begin{table}[h]
\centering
\renewcommand{\arraystretch}{1.2}
\setlength{\tabcolsep}{4pt}
\small
\begin{tabular}{lcccc}
\toprule
& \multicolumn{3}{c}{Per-halo boost $e^{\sigxiline^2(\Mh)}$} & Mass-int.\\
\cmidrule(lr){2-4}
Line ($t_\lambda$) & $10^9\Msun$ & $10^{10}\Msun$ & $10^{11}\Msun$ & $\mathcal{B}_\lambda$\\
\midrule
$\Halpha$  ($7$~Myr)         & $63$  & $18$  & $6.2$ & $6.8$ \\
$\CII$     ($50$~Myr)        & $22$  & $7.7$ & $3.5$ & $3.6$ \\
CO(1--0)   ($80$~Myr)        & $15$  & $5.7$ & $2.8$ & $2.9$ \\
UV cont.   ($100$~Myr)       & $13$  & $4.9$ & $2.5$ & $2.6$ \\
\bottomrule
\end{tabular}
\caption{\textbf{Per-halo and mass-integrated shot-noise boosts} at $z\!\approx\!6$ for
the M26 central burstiness parameters $\sigPS(10^{11}\Msun)\!=\!2.0$, $d\sigPS/d\log_{10}\Mh\!=\!-0.5$, $\tauPS\!=\!25\Myr$, computed from
Eq.~\eqref{eq:Vexact}.
$\mathcal{B}_\lambda$ uses the Sheth--Tormen~\cite{ShethTormen:1999} HMF via
\texttt{oLIMpus}~\cite{Libanore:2025}/\texttt{Zeus21}~\cite{Munoz:2023Zeus}
(Planck 2018 cosmology, \texttt{CLASS}~\cite{Lesgourgues:2011}) at $z=6$, with
an M26-style double power-law SFE for $\Lbar_\lambda(\Mh)$.
The dominant systematic is the assumed $\sigPS$, displayed explicitly in Fig.~\ref{fig:robustness}.
}
\label{tab:boost}
\end{table}

For the M26 central values, the mass-integrated boosts at $z\!\approx\!6$ are
$\mathcal{B}_{\Halpha}\!\simeq\!6.8$, $\mathcal{B}_{\CII}\!\simeq\!3.6$,
$\mathcal{B}_{\rm CO}\!\simeq\!2.9$, $\mathcal{B}_{\rm UV}\!\simeq\!2.6$.
These are substantially larger than the $\sigma_L\!\sim\!0.3$~dex phenomenological
scatter routinely assumed in LIM forecasts~\cite{Sun:2019,Schaan:2021} (which yields a
line-independent boost $e^{(\sigma_L\ln 10)^2}\!\simeq\!1.6$), for two reasons: 
(i) $\sigPS$ inferred by M26 is much larger than $0.3$~dex, and (ii) the mass dependence
weights the integral toward lower-mass halos where $\sigPS$ is even larger. A third effect, the
SPS Green's function convolution that defines $V_\lambda$,  pulls the boost down for long-window lines relative to a naive 
$e^{\sigma_x^2}$  prescription using the M26 amplitude (Appendix~B).
Substituting phenomenological line-specific $\Lbar_\lambda(\Mh)$ shapes for $\Halpha$, [O\,\textsc{iii}], $\CII$, and CO into the
boost integral shifts $\mathcal{B}_\lambda$ at the $-35\%$ to $+17\%$ level relative to the
$\Lbar_\lambda\!\propto\!\meanSFR$ baseline (Appendix~A).

\begin{figure}[t]
\centering
\includegraphics[width=0.925\columnwidth]{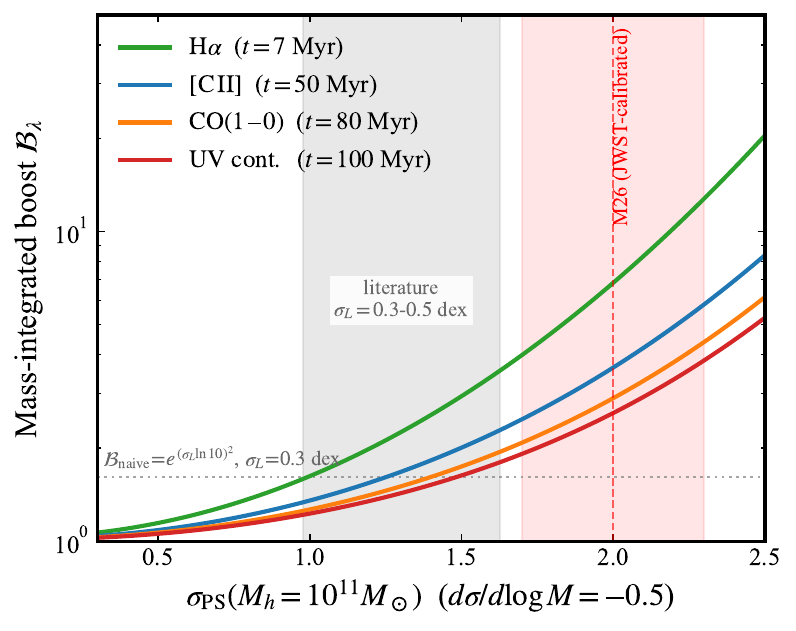}
\caption{\textbf{Sensitivity of the mass-integrated shot-noise boost
$\mathcal{B}_\lambda$ to the assumed burstiness amplitude $\sigPS$,}
at $z\!=\!6$ with $\tauPS\!=\!25$~Myr and the
M26 mass-dependent slope $d\sigPS/d\log_{10}\Mh\!=\!-0.5$, plotted
against the value at the pivot mass $\Mh\!=\!10^{11}\Msun$.  Red
shaded band marks the M26 1$\sigma$ region $\sigPS\!=\!2.0\pm 0.3$; gray band marks the literature 
$\sigma_L\!=\!0.3$--$0.5$ dex region (translated to our $\sigPS$ via $\sigPS\!=\!\sqrt{2}\ln(10)\sigma_L$).
Dotted horizontal line marks the line- and mass-independent boost $e^{(\sigma_L\ln 10)^2}\!\simeq\!1.6$ obtained
under the literature $\sigma_L\!=\!0.3$~dex prescription, lower by factors of $\sim\!2$--$5$ than our  $\mathcal{B}_\lambda$ from M26 --- a gap that translates directly into the gray dash-dot vs solid green offset in the absolute-units forecast of Fig.~\ref{fig:pk}.
}
\vspace{-0.15in}
\label{fig:robustness}
\end{figure}

\smallskip
\noindent\textit{Sensitivity to $\sigPS$.}~%
Because the boost depends exponentially on $\sigma_x^2\!=\!\sigPS^2/2$,
its absolute value is sensitive to the assumed burstiness amplitude.
Fig.~\ref{fig:robustness} shows $\mathcal{B}_\lambda$ vs.\ $\sigPS$ at the M26 mass-dependent slope, with the literature $\sigma_L\!=\!0.3$--$0.5$~dex region marked for reference.  
For $\sigPS\!\lesssim\!1$ the boost is modest, $\mathcal{B}_\lambda\!\lesssim\!2$.  At the M26 central value $\sigPS\!\simeq\!2$, $\mathcal{B}_\lambda$ ranges from $\sim\!2.6$ for UV to $\sim\!6.8$ for $\Halpha$, factors of $\sim\!2$--$5$ above the literature $\sigma_L\!=\!0.3$~dex prescription. A measurement of $\Pshot^\lambda$, given external estimates of $\Lbar_\lambda(\Mh)$, provides a direct route to $\sigPS$ from LIM alone.

\begin{figure}[t]
\centering
\vspace{-0.1in}
\includegraphics[width=0.925\columnwidth]{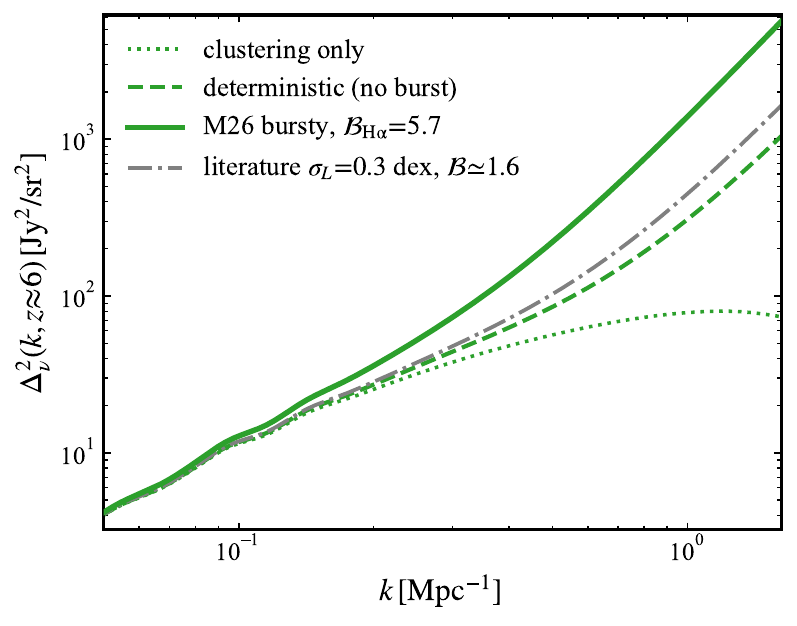}
\vspace{-0.05in}
\caption{\textbf{$\Halpha$ LIM auto-power spectrum at $z\!\approx\!6$, with and without M26 burstiness.}  Solid: full $\Delta^2_\nu(k)$ including the M26-boosted shot noise ($\mathcal{B}_{\Halpha}\!\simeq\!5.7$).  Dashed: deterministic prediction (no burstiness).  Dotted: clustering component, unchanged between cases.  Gray dash-dot: literature $\sigma_L\!=\!0.3$~dex prescription, $\mathcal{B}\!\simeq\!1.6$ --- the standard pre-M26 LIM scatter ansatz, applied as a multiplicative correction to the same Yang+24 deterministic shot noise for visual reference; partition issues are discussed in Appendix~A.  $\Halpha$ uses the Yang+24 phenomenological $\Lbar_{\Halpha}(\Mh)$ shape with the M26 SFE substituted in for the underlying mean SFR; mean intensity, effective bias, and shot-noise normalization are taken directly from \texttt{oLIMpus}~\cite{Libanore:2025}.  Both bursty and deterministic spectra are anchored to the same UVLF-calibrated mean SFR; they describe the same physical universe with $\langle\dot M_\star\rangle$ matched to JWST data, differing only in whether the SFR scatter around that mean is at its M26 best-fit ($\sigPS\!=\!2.0$, $\tauPS\!=\!25$~Myr) or zero.  The boost factor here is slightly smaller than the $\Lbar\!\propto\!\meanSFR$ baseline value of $6.8$ in Tab.~\ref{tab:boost}, with the difference dictated by the Yang+24 shape; cf.\ Tab.~\ref{tab:appendix_boost}.  See Fig.~\ref{fig:appendix_pk} (Appendix~A) for the analogous figure at the survey-relevant redshifts of $\CII$ ($z\!=\!6$, TIME/EXCLAIM/CCAT) and CO(1--0) ($z\!=\!3$, COMAP).}
\vspace{-0.05in}
\label{fig:pk}
\end{figure}

\smallskip
\noindent\textit{Observable impact.}~%
The full impact of the boost on observables is shown in Fig.~\ref{fig:pk}: the $\Halpha$ LIM auto-power spectrum at $z\!=\!6$, with and without M26 burstiness, computed under the Yang+24 phenomenology with the M26 SFE substituted in for self-consistency.  Burstiness leaves the clustering term unchanged in our convention; at large $k$ the shot-noise tail is amplified by $\mathcal{B}_{\Halpha}\!\simeq\!5.7$, lifting the bursty curve nearly an order of magnitude above its deterministic counterpart at $k\!\gtrsim\!1$~Mpc$^{-1}$, a striking change. 

For survey-specific prospects (cf.\ Appendix~A for line-modeling details), the M26 central values imply $\mathcal{B}\!\sim\!4$--$10$ for $\Halpha$ at $z\!\sim\!3$--$8$, potentially detectable by SPHEREx~\cite{SPHEREx:2014} at the lower-$z$ end of its sensitivity range, where $\Halpha$ sits well within its primary spectral coverage. For $\mathcal{B}_{\CII}\!\sim\!2$--$6$ at $z\!\sim\!4$--$8$, EXCLAIM~\cite{Cataldo:2021} and CCAT/FYST~\cite{CCAT:2022} are well-positioned: forecasts for  FYST $\CII$ find detectability of the $\CII$ power spectrum out to $z\!\approx\!5.8$ (and potentially $z\!\approx\!7.4$) depending on the adopted signal model~\cite{Karoumpis:2022}, placing the $z\!=\!6$ burst-enhanced signal (Appendix~A) squarely in the detectable regime. For CO(1--0), $\mathcal{B}_{\rm CO}\!\sim\!2$ for COMAP-EoR~\cite{Breysse:2022} at $z\!\sim\!5$--$8$; at $z\!=\!3$, the lower-$z$ main sequence drives the kernel-peak halo mass to $\sim\!10^{12}\Msun$ where $\sigPS$ is much smaller and the long Yang+21 SPS window further suppresses the boost to $\mathcal{B}_{\rm CO}\!\simeq\!1.6$. The implied bursty shot-noise power is well below the COMAP Season~2 upper limit~\cite{Stutzer:2024} but may be accessible to the full 5-year campaign (pre-Season 2 forecast S/N $9$--$17$~\cite{Chung:2022}). 

It is important to caution that the redshift evolution above, and the redshift-tomography prediction below, assume the M26 amplitude and slope are themselves $z$-independent. This working assumption is not directly constrained by M26's $z\!\sim\!4$--$6$ calibration; a measured deviation from these predictions would itself be diagnostic of $z$-evolution of the burstiness parameters. Quantifying detectability in specific configurations will require dedicated Fisher forecasts. Beyond this raw-amplitude effect, the spectral and tomographic structure of the boost provides two additional diagnostics of the underlying burstiness, which we now turn to.

\bigskip
\noindent\textit{Cross shot noise between lines.}~%
For two lines $\lambda_{1,2}$ at the same redshift, the cross-power
spectrum is
$P_{\lambda_1\lambda_2}(k)\!=\!\Ibar_1\Ibar_2\,\beff(\lambda_1)\beff(\lambda_2)\Pm(k)+\Pshot^{\lambda_1\lambda_2}$,
with the same effective biases as in Eq.~\eqref{eq:Pk_decomp}.
Since different halos are uncorrelated, the cross shot noise is
\be
\Pshot^{\lambda_1\lambda_2}\!=\!\mathcal{X}_1\mathcal{X}_2\!
\int\!d\Mh\,\dn\,\langle L_1 L_2\rangle(\Mh),
\ee
with the same-halo cross moment given by~\cite{Munoz:2026}
\be
\langle L_1 L_2\rangle - \Lbar_1\Lbar_2 =
\int\!\frac{d\omega}{2\pi}\,\tilde W_1^*(\omega)\tilde W_2(\omega)\,P_y(\omega)\,.
\label{eq:CrossVar}
\ee
This is the Fourier-space overlap of the two window functions weighted
by the SFR power spectrum.  Defining
$\sigma_{\xi,12}^2(\Mh)\!\equiv\!\ln[1\!+\!{\rm cov}(L_1,L_2)/(\Lbar_1\Lbar_2)]$, we get
\be
\Pshot^{\lambda_1\lambda_2} = \mathcal{X}_1\mathcal{X}_2\!
\int\!d\Mh\,\dn\,\Lbar_1\Lbar_2\,e^{\sigma_{\xi,12}^2(\Mh)}\,.\;
\label{eq:Pshot_cross}
\ee

\noindent\textit{Cross-correlation coefficient.}~To isolate the burst temporal structure from the deterministic luminosity relations, we construct a fractional-excess estimator. With $\Delta_{\lambda_1\lambda_2}\!\equiv\!(P_{\rm shot}^{\lambda_1\lambda_2}/P_{\rm shot}^{\lambda_1\lambda_2}|_{\rm det})\!-\!1$ (and analogously for the autospectra), we define
\begin{equation}
\mathcal{R}_{\lambda_1\lambda_2} \equiv \frac{\Delta_{\lambda_1\lambda_2}}{\sqrt{\Delta_{\lambda_1} \Delta_{\lambda_2}}}.
\label{eq:R}
\end{equation}
By construction, $\mathcal{R}$ is the correlation coefficient of the \emph{bursty excess} alone: the deterministic shot noise is divided out in numerator and denominator, leaving only the burstiness-induced contribution. The overall normalization of $\Lbar_\lambda$ cancels (only the shape of the L--$\Mh$ relation matters), and in the limit of mass-independent $\sigPS$ the HMF cancels exactly, reducing $\mathcal{R}$ to the per-halo Pearson correlation $V_{12}/\sqrt{V_1 V_2}$.  In the M26 mass-dependent case it acquires a mild $M_h$-weighting. The asymptotic limits are $\mathcal{R}\!\to\!1$ at $\tauPS\!\to\!\infty$ (bursts coherent over arbitrarily long times, both windows seeing the same burst) and $\mathcal{R}\!\to\!\sqrt{\min(t_{\lambda_1},t_{\lambda_2})/\max(t_{\lambda_1},t_{\lambda_2})}$ at $\tauPS\!\to\!0$ (geometric window overlap); the M26-favored regime $\tauPS\!\sim\!10$--$50$~Myr lies in between, where the gradient with $\tauPS$ is steepest. As a population-level analog of the resolved-galaxy $\Halpha$/UV diagnostic of M26, $\mathcal{R}$ probes $\tauPS$ at redshifts where individual galaxies are too faint to be detected.

\begin{figure*}[t]
\centering
\includegraphics[width=0.92\textwidth]{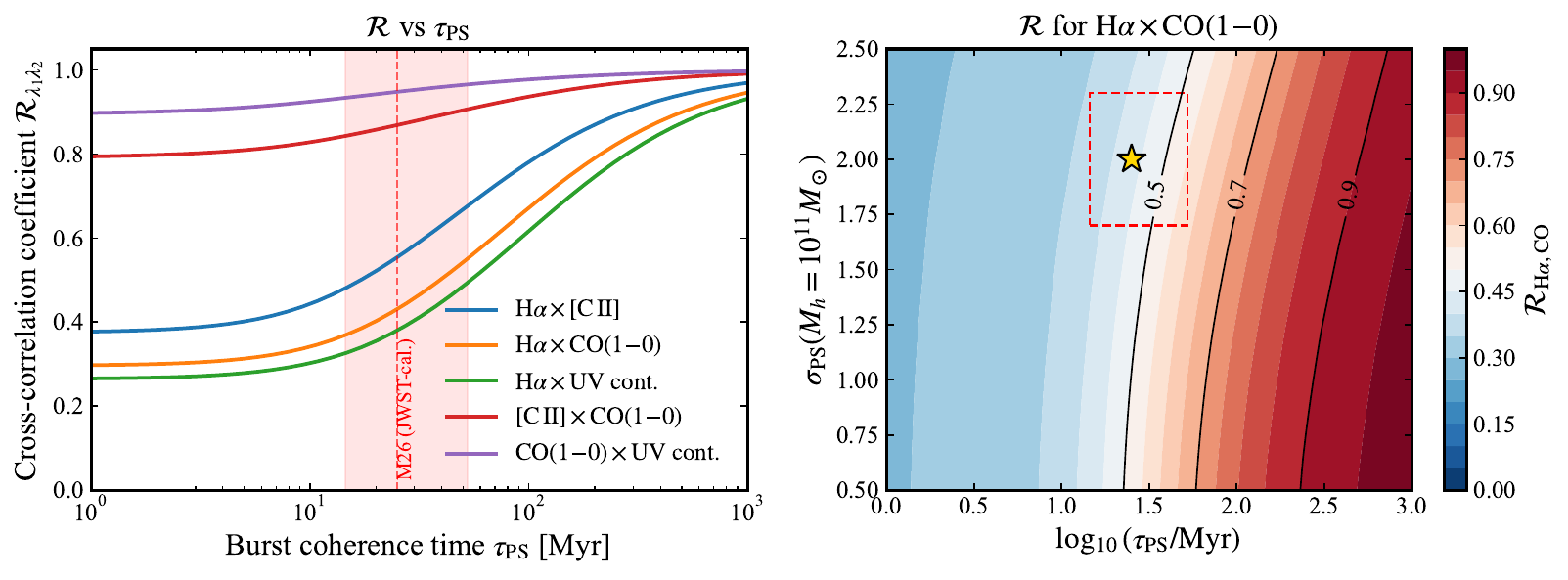}
\caption{\textbf{Line cross-correlation coefficient
$\mathcal{R}_{\lambda_1\lambda_2}$ as a probe of the burst coherence
time $\tauPS$.}  $\mathcal{R}$ is defined per Eq.~\eqref{eq:R} and isolates the bursty shot noise excess (by removing the deterministic luminosity weighting, leaving only the correlated stochastic excess).
{\it Left:} $\mathcal{R}$ vs.\ $\tauPS$ for five line pairs.
Pairs spanning very different temporal scales ($\Halpha\!\times\!$UV, $\Halpha\!\times\!$CO) carry the most $\tauPS$ information; pairs with similar windows (CO$\,\times\,$UV) saturate near unity and serve as
cross-checks. Vertical dashed line and red band mark the M26 1$\sigma$ region for $\tauPS$.
{\it Right:} $\mathcal{R}_{\Halpha,\rm CO}$ over the M26 $(\sigPS(M_h^{\rm pivot}),\tauPS)$ plane.  Near-vertical isocontours show $\mathcal{R}$ depends primarily on $\tauPS$, only weakly on $\sigPS(M_h^{\rm pivot})$.  Yellow star: M26 central values; red dashed rectangle: M26 1$\sigma$ region.}
\label{fig:R}
\end{figure*}

Fig.~\ref{fig:R} shows $\mathcal{R}_{\lambda_1\lambda_2}(\tauPS)$ for five line pairs, computed with the \texttt{oLIMpus}~\cite{Libanore:2025} HMF and M26 SFE.  Pairs spanning very different timescales (e.g.\ $\Halpha\times$UV, $\Halpha\times$CO) carry the most $\tauPS$ information, interpolating between the short- and long-burst limits across the M26-favored band; pairs with similar windows (CO$\times$UV) saturate near unity and serve as cross-checks.
At the M26 best fit, $\mathcal{R}_{\Halpha\times{\rm CO}}\!\approx\!0.5$, between the long-burst ($\mathcal{R}\!\to\!1$) and short-burst ($\mathcal{R}\!\to\!0.3$) limits and providing a clean handle on $\tauPS$.  The right panel maps $\mathcal{R}_{\Halpha,\rm CO}$ over the M26 $(\sigPS,\tauPS)$ plane: isocontours are nearly vertical, confirming that $\mathcal{R}$ depends primarily on $\tauPS$ and only weakly on $\sigPS$. Within the M26 SFR-driven framework, $\mathcal{R}\!\ge\!0$ follows from $\xi_x\!\ge\!0$ and $W_\lambda\!\ge\!0$; a measurement of $\mathcal{R}\!<\!0$ would diagnose physics beyond M26 (e.g., SFR-correlated dust attenuation or feedback-mediated quenching).

\begin{figure}[t]
\centering
\vspace{-0.15in}
\includegraphics[width=0.925\columnwidth]{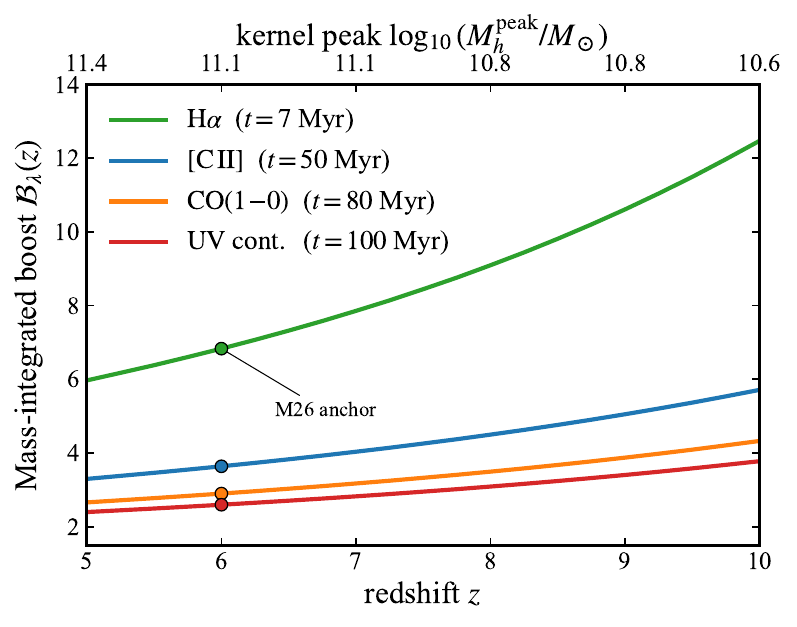}
\caption{\textbf{Redshift tomography of   shot-noise boost.}
$\mathcal{B}_\lambda(z)$ for our three representative lines $\Halpha$,
$\CII$, and CO(1--0), computed
using  \texttt{oLIMpus}~\cite{Libanore:2025} and the M26 mass-dependent
$\sigPS(\Mh)$, with all other parameters held fixed.  Markers
indicate the M26 anchor at $z\!=\!6$.  Top axis: the kernel-peak halo
mass $\log_{10}(\Mh^\mathrm{peak}/\Msun)$ at each redshift, computed
from the $\Lbar_\lambda^2(dn/d\ln \Mh)$ weighting.  As the HMF cutoff
$M_*(z)$ falls with redshift, lower-mass halos with larger $\sigPS$
dominate the integral, and $\mathcal{B}_\lambda$ rises steeply.  The
$\Halpha$ boost grows from $\mathcal{B}_{\Halpha}\!\simeq\!6.8$ at
$z\!=\!6$ to $\mathcal{B}_{\Halpha}\!\simeq\!12.5$ at $z\!=\!10$.
Extrapolation to $z\!\sim\!10$ is subject to caveats (see the text), including faint-end SFE evolution and breakdown of the
lognormal description for $\sigPS\!\gtrsim\!3$.}
\vspace{-0.05in}
\label{fig:redshift}
\end{figure}

\smallskip
\noindent\textit{Redshift tomography.}~%
A complementary and arguably the most direct strategy for using LIM as a quantitative probe of mass-dependent burstiness is to observe the same emission line across multiple redshift bins.  Because the $\Lbar_\lambda^2(dn/d\ln \Mh)$ kernel that drives shot noise tracks the HMF cutoff $M_*(z)$, the dominant halo mass shifts from $\Mh\!\simeq\!10^{11.1}\Msun$ at $z\!=\!6$ to $\simeq\!10^{10.6}\Msun$ at $z\!=\!10$.  Under the M26 slope $d\sigPS/d\log_{10}\Mh\!=\!-0.5$, the kernel-peak $\sigPS$ grows from $\simeq\!2.0$ at $z\!=\!6$ to $\simeq\!2.2$ at $z\!=\!10$, and $\mathcal{B}_\lambda(z)$ grows correspondingly: a factor of $\sim\!2$ for $\Halpha$ between $z\!=\!6$ and $z\!=\!10$ (Fig.~\ref{fig:redshift}). Since the prediction depends only on the M26 slope and HMF evolution (per-line normalizations of $\Lbar_\lambda$ cancel in the boost ratio), it constitutes a direct test of M26's mass-dependent burstiness picture. The main caveats to this are that the faint-end SFE evolution can shift $\mathcal{B}_\lambda(z)$ at the $\mathcal{O}(10\%)$ level, and that at $z\!\gtrsim\!10$ the kernel of lines with extended low-mass $\Lbar_\lambda(\Mh)$ shapes can descend to halo masses where the implied $\sigPS\!\gtrsim\!3$ approaches the regime in which gas depletion and stellar feedback regulate the SFR and the lognormal description of its scatter likely breaks down. The qualitative finding that $\mathcal{B}_\lambda$ rises with $z$ is robust to both effects.

\smallskip
\noindent\textit{Implications.}~%
What does an inflated shot noise mean for actual LIM experiments? In the shot-noise-dominated regime, $\Pshot^\lambda$ is itself a measurable signal, and one that, once the deterministic prediction is known, carries information about the underlying burstiness.  Its amplitude, given external estimates of $\Ibar_\lambda$ and $\beff$, constrains $\mathcal{B}_\lambda$ and hence $\sigPS$, with degeneracies against the $\Lbar_\lambda(\Mh)$ relation, halo duty cycles, and selection effects partially broken by the same-redshift line ratio $\Pshot^{\lambda_1}/\Pshot^{\lambda_2}$.  The cleanest probe of the \emph{mass dependence} of $\sigPS$ is the redshift tomography of Fig.~\ref{fig:redshift}: observing a single line across multiple $z$ slides the LIM kernel in $\Mh$ in a calculable way and  maps $\mathcal{B}_\lambda(z)$ onto $\sigPS(\Mh)$ with the line itself as the calibrating common reference.  Complementing this, the cross-line coefficient $\mathcal{R}_{\lambda_1\lambda_2}$ provides a handle on $\tauPS$ specifically (Fig.~\ref{fig:R}), since it depends primarily on the relative SPS window widths.  A further diagnostic is the LIM clustering-to-shot ratio, suggested as a means to reconstruct the LF~\cite{YueFerrara:2019}: this ratio is itself biased by $\mathcal{B}_\lambda$, so naive reconstruction yields a biased result, but with independent UVLF input the same ratio inverts into a measurement of $\mathcal{B}_\lambda$ and hence $\sigPS$.

Mass-dependent burstiness brings a structural benefit for \emph{interloper-line contamination}.  Both the high-$z$ target and any lower-$z$ interloper line are boosted, but lower-$z$ interlopers originate from more massive halos with smaller $\sigPS$ and are therefore  less boosted than the target.  The net effect is to raise the target-to-interloper signal ratio in the shot-noise tail, partially mitigating one of the main systematic concerns in high-$z$ LIM. The size of this benefit depends on the redshift contrast and the mass scales involved and is best quantified survey-by-survey.

The shot-noise enhancement is, in the conventional reading, bad news for cosmology: in the clustering-dominated regime, an inflated stochastic floor lowers the effective number of accessible Fourier modes, shifting the scale at which boosted shot noise crosses the clustering term to larger physical scales and reducing the range of $k$ over which features of $P_m(k)$ such as the baryon acoustic oscillations (BAOs) remain accessible above the floor; this can degrade BAO and expansion-rate measurements from future LIM surveys~\cite{Bernal:2019gfq,Bernal:2019jdo}. The flip side is more attractive: the enhanced shot noise is itself a measurement, and an inversion of the problem  recasts the LIM shot noise, through its $z$-tomography and cross-line ratio $\mathcal{R}$, as a probe of $\sigPS(\Mh,z)$ and $\tauPS$ at redshifts and halo masses inaccessible to resolved-galaxy spectroscopy. 

\smallskip
\noindent\textit{Discussion.}~%
Several simplifications underlie our results.  At fixed $\Mh$ we have taken $L_\lambda$ lognormal, assumed uniform SPS Green's functions across the galaxy population, and have not modeled bright-source masking or flux cuts.  All these are straightforward refinements left to experimental forecasts. A few assumptions warrant further comment.

\vspace{2pt}
\emph{The OU correlator.}~The exponential form $\xi_x(\Delta t)\!=\!\sigma_x^2 e^{-|\Delta t|/\tauPS}$ inherited from M26 has a non-differentiable cusp at $\Delta t\!=\!0$, unphysical at the few-Myr timescale of cloud collapse and feedback.  Substituting a smoother Gaussian kernel $\xi_x\!\propto\!e^{-\Delta t^2/(2\tauPS^2)}$ at fixed $\tauPS$ increases the per-halo $V_\lambda$ by $\sim\!20\%$ for $\Halpha$ and up to $\sim\!90\%$ for UV, owing to the Gaussian's larger integrated coherence ($\tauPS\sqrt{\pi/2}$ vs.\ $\tauPS$ for OU); matching the integrated coherence roughly halves these shifts. After mass integration the residual systematic on $\mathcal{B}_\lambda$ is $\sim\!10\%$ for $\Halpha$ and up to $\sim\!50\%$ for the long-window tracers ($\CII$, CO, UV). Crucially, these shifts are \emph{upward}: under any plausible alternative to OU, our boost numbers grow rather than shrink, so the conclusion that $\mathcal{B}_\lambda$ exceeds the literature $\sigma_L\!=\!0.3$~dex value of $\sim\!1.6$ is robust against kernel choice.

\vspace{2pt}
\emph{Inter-halo SFR correlations.}~We have assumed halo-to-halo independence of $x(t)$.  If burstiness is environmentally triggered (by mergers, filamentary accretion, large-scale assembly bias, reionization fronts, etc.), neighboring halos can have correlated fluctuations.  This appears as a \emph{scale-dependent} modification of $\beff$ at large scales, set by the correlation length of the trigger (negligible for merger-driven bursts of length corresponding to virial radii but potentially $\sim\!$~Mpc for accretion-driven bursts), together with a two-halo contribution to the shot-noise term. Both contributions are super-Poisson and would bias the inferred $\mathcal{B}_\lambda$ \emph{upward} from the per-halo value computed here. We defer quantitative estimates of these to future work.

\vspace{2pt}
\emph{Validity of the lognormal approximation.}~The shot-noise expression Eq.~\eqref{eq:Pshot_main} depends only on the second moment of $L_\lambda$, which Eq.~\eqref{eq:Vexact} computes \emph{exactly} given the OU correlator; the lognormal interpretation merely labels the result as $e^{\sigxiline^2}$. M26 verify (their Appendix~D) that the per-halo lognormal PDF for UV agrees with direct simulation at the $\sim\!10\%$ level, sufficient for our applications. Statistics beyond the second moment, such as the voxel intensity distribution~\cite{Breysse:2017} or the line-intensity bispectrum~\cite{MoradinezhadDizgah:2019}, depend on higher cumulants of the SFR field and require the M26 short/long-timescale decomposition rather than the effective single-timescale framework adopted here.

\vspace{2pt}
\emph{Single-timescale assumption.}~%
Our OU autocorrelator $\xi_x(\Delta t)\!=\!\sigma_x^2 e^{-|\Delta t|/\tauPS}$
assumes a single coherence time, while real galaxy star formation histories vary on a range of
physical timescales: rapid fluctuations from individual star-forming regions,
intermediate cycles driven by stellar feedback, and slower modulation from
gas accretion and mergers~\cite{Iyer:2020}. For windows $t_\lambda \gg \tauPS$
(UV, CO), the formalism correctly averages over many coherence times, but
any long-timescale ($\gtrsim\!100$~Myr) SFR variability beyond what M26's
$\Halpha$/UV calibration constrains would add variance not captured here,
modestly increasing $\mathcal{B}_\lambda$ for the longer-window tracers.
The $\Halpha$ window is short enough ($t_\lambda\!\sim\!\tauPS/4$) that this
concern does not apply.

\vspace{2pt}
\emph{Other sources of scatter.}~Our framework attributes all luminosity scatter at fixed halo mass to the M26 OU burstiness process. Additional non-burstiness scatter, from dust attenuation correlated with SFR, geometric/orientation effects, central--satellite distributions, or resonant radiative transfer for lines like $\Lya$~\cite{Silva:2013,Heneka:2021aey}, would add in quadrature to $\sigma_x^2$ in our formalism; for non-burstiness scatter at the $\sim\!0.1$--$0.2$~dex level the additional enhancement is at the $\sim\!20$--$30\%$ level.

\smallskip
\noindent\textit{Conclusions.}~%
The picture of high-redshift star formation emerging from \emph{JWST}, of large, time-correlated SFR scatter especially in the low-mass halos that dominate LIM signals, forces us to revisit predictions for the LIM shot noise.  
Adopting the M26 burstiness framework,  we  developed a formalism to consistently account for burstiness in LIM measurements and calculate how much the shot noise contribution is enhanced over the deterministic contribution from halo discreteness, as a function of halo mass. 

For the M26 central values, the  total, mass-integrated boosts at $z\!\sim\!6$ are $\mathcal{B}\!\sim\!2.5$--$7$ across our list of reference lines, growing to $\mathcal{B}\!\simeq\!12.5$ for $\Halpha$ at $z\!\sim\!10$ via redshift tomography. The boost depends exponentially on $\sigPS$ and falls back to $\mathcal{B}\!\lesssim\!2$ if $\sigPS\!\lesssim\!1$ (Fig.~\ref{fig:robustness}). These are factors of $2-5$ larger than the typically adopted values for scatter in LIM forecasts, with positive implications for detection.  

The shot-noise enhancement also has consequences for cosmology: a higher stochastic floor obscures BAO features and degrades expansion-rate measurements~\cite{Bernal:2019gfq, Bernal:2019jdo}.  Conversely and more attractively, the LIM shot noise itself, in particular its $z$-tomography and cross-line ratio $\mathcal{R}$, can be inverted to constrain $\sigPS$ and $\tauPS$ at redshifts and halo masses inaccessible to resolved-galaxy spectroscopy. 

Forthcoming work will integrate our results into Fisher forecasts for COMAP, EXCLAIM, SPHEREx, and FYST, and extend the formalism to higher-order statistics~\cite{Finish:2026}. Another work will focus on the bursty 21cm signal, where the  effect is both more subtle and more prominent~\cite{Hovav:2026}.

\vspace{-0.1in}
\section*{Acknowledgments}
\vspace{-0.1in}
E.D.K.\ acknowledges support from the U.S.--Israel
Binational Science Foundation (NSF-BSF grant 2022743 and BSF grant
2024193) and the Israel National Science Foundation (ISF grant 3135/25),
as well as from the joint Israel--China program (ISF--NSFC grant
3156/23). H.~L.\ is supported by a Negev PhD fellowship from the Kreitman school at BGU. E.~V.\ is supported by an Azrieli International postdoctoral fellowship at TAU. J.B.M.\ acknowledges support from NSF Grants AST-2307354 and
AST-2408637, and the NASA grant JWST-GO-03224.

\bibliographystyle{plain}

\appendix

\section*{Appendix A: Line-modeling considerations}

The body of this paper assumes the minimal ansatz
$\Lbar_\lambda(\Mh)\!\propto\!\meanSFR(\Mh)$, so that the boost factor
$\mathcal{B}_\lambda$ depends only on the temporal kernel $t_\lambda$ and
the M26 burstiness parameters $(\sigPS, \tauPS)$.  Real LIM line models
are richer than this: they invoke metallicity-, ISM-, and feedback-dependent
modifications of the $\Lbar_\lambda(\Mh)$ shape, and some are calibrated
against data that already contain galaxy-to-galaxy scatter.  In this
appendix we briefly classify the different kinds of line models in current
use, quantify the resulting shifts in $\mathcal{B}_\lambda$ when their
$\Lbar_\lambda(\Mh)$ shapes are substituted into our boost integral, and
flag a partition issue that arises when our boost is combined with
luminosity-function-anchored prescriptions.  We do not attempt a complete
forecast revision here; that requires line-by-line and survey-by-survey
treatment beyond the current scope.

\smallskip
\noindent\textit{Three classes of line-luminosity prescriptions.}~%
LIM forecasts in the literature fall into three broad categories that
interact differently with our boost:

\noindent(i) \emph{Mean-SFR-anchored.}  $L_\lambda(\Mh) = \kappa_\lambda\,\meanSFR(\Mh)$
with $\meanSFR$ tied to a UVLF-calibrated SFE.  This is the framework
adopted in the body of this work and in modern semi-numerical LIM
codes~\cite{Munoz:2023Zeus,Libanore:2025}.  The boost applies cleanly:
SFR scatter at fixed $\Mh$ is exactly what M26 quantifies, and the SPS
Green's function carries it to $L_\lambda$.

\noindent(ii) \emph{Empirical $L(\Mh)$ relations.}  Power-law or
double-power-law fits to the population-level $L_\lambda$--$\Mh$ relation,
from semi-analytic galaxy formation
models~\cite{Lagache:2018} or directly from
simulations~\cite{Yue:2015,YueFerrara:2019}.  These prescribe a
deterministic $L_\lambda(\Mh)$ shape with no explicit scatter; the
implicit assumption is that the calibration data are well-described by
their mean.  Applying our boost on top is in principle clean, with one
caveat: when the calibration is performed on a population that includes
bursty galaxies, the dispersion is partially absorbed into the slope of
the fit.  Whether this constitutes a double-count depends on whether the
calibration sample is representative of the M26 high-burstiness regime and left to the user to resolve.

\noindent(iii) \emph{Direct LF-based forecasts.}  Codes that integrate over an
observed line luminosity function (LF) to predict shot
noise~\cite{Li:2016,Breysse:2014,Yue:2015,Sun:2019,SPHEREx:2014,Cataldo:2021,COMAP:2022,CCAT:2022}.
The observed LF contains the physical scatter automatically. Therfore, multiplying
their predicted $\Pshot^\lambda$ by our $\mathcal{B}_\lambda$ re-anchors 
to the M26 burstiness inference and is informative as a
consistency check, but a full predictive forecast would require modeling
how the observed LF is itself generated by burstiness.

The phenomenological $L(\Mh)$ models in class (ii) give the cleanest
test of how line-specific $\Lbar_\lambda(\Mh)$ shapes shift our boost,
and we use them below.

\begin{figure}[t]
\centering
\includegraphics[width=0.96\columnwidth]{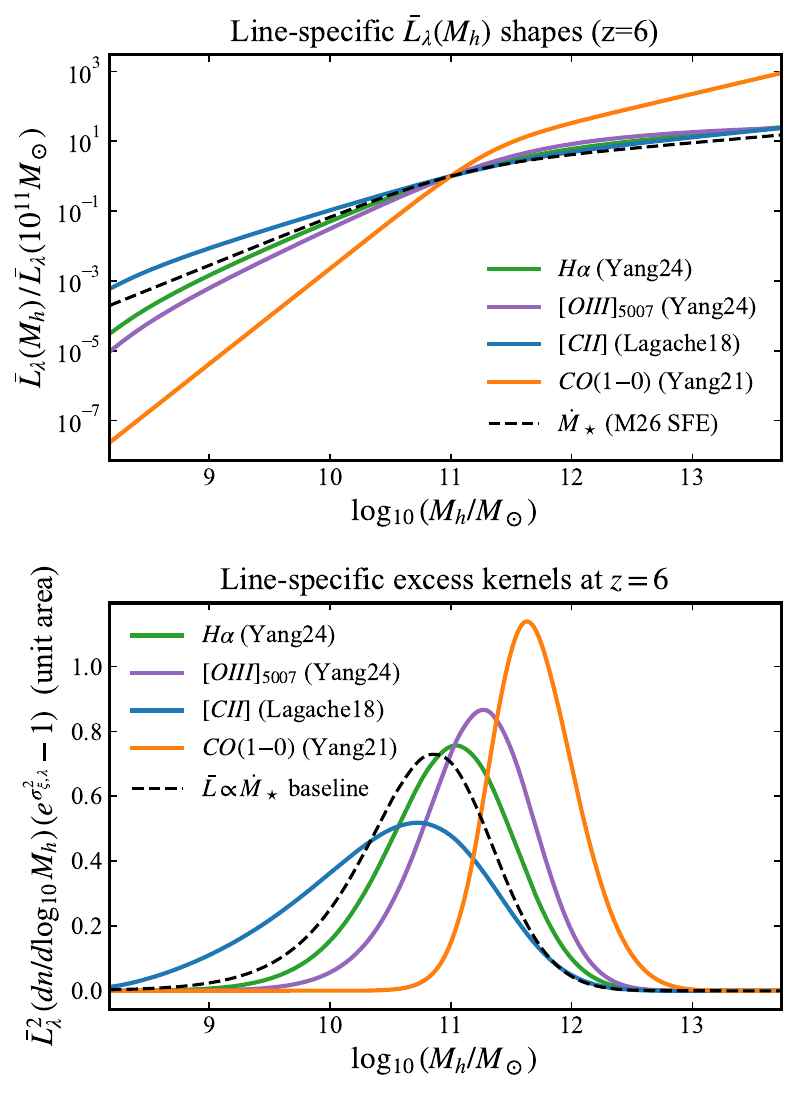}
\caption{\textbf{Line-specific shape modifications at $z=6$.}
{\it Top:} $\Lbar_\lambda(\Mh)$ for four phenomenological line models,
each normalized to its value at $\Mh\!=\!10^{11}\Msun$ and compared to the
mean SFR shape (dashed black, M26 SFE).  The $\CII$ Lagache+18 fit is
sublinear at low mass; the Yang+21 CO(1--0) prescription is the most
distinctive, with a steep low-mass cutoff and a super-linear high-mass
slope, both reflecting metallicity- and shielding-dependent CO formation
that is sharply truncated below a critical halo mass.
{\it Bottom:} area-normalized excess kernels under each line-specific shape.
Compared to the body's $\Lbar_\lambda\!\propto\!\meanSFR$ kernels, the
peak of the CO excess shifts to higher $\Mh$ (driven by the steep low-mass
CO cutoff), while the $\CII$ excess broadens to lower $\Mh$ (driven by
the sublinear slope).}
\label{fig:appendix_kernels}
\end{figure}

\smallskip
\noindent\textit{Line-specific shape modifications.}~%
We substitute the M26 SFE shape into \texttt{oLIMpus} and run four
representative phenomenological line models:
$\Halpha$ and $[$O\,\textsc{iii}$]_{5007}$ (Yang+24),
$\CII$ (Lagache+18~\cite{Lagache:2018}), and CO(1--0) (Yang+21,
extrapolated from the published $J\!=\!2\!-\!1$ table).  This isolates the
effect of the line-specific $\Lbar_\lambda(\Mh)$ shape on the boost, with
all other physics (HMF, $\sigPS(\Mh)$, $\tauPS$, SPS window) held fixed
to the body's values.
Fig.~\ref{fig:appendix_kernels} shows the resulting $\Lbar_\lambda(\Mh)$
shapes (top, normalized at the M26 pivot $10^{11}\Msun$) and the area-normalized
excess kernels $\Lbar_\lambda^2 (dn/d\log_{10}\Mh)(e^{\sigxiline^2}-1)$ at $z=6$
(bottom).

\begin{table}[h]
\centering
\renewcommand{\arraystretch}{1.15}
\setlength{\tabcolsep}{6pt}
\small
\begin{tabular}{lcccc}
\toprule
Line & $t_\lambda$ & $\mathcal{B}_\lambda$ & $\mathcal{B}_\lambda$ & $\Delta$ \\
     & [Myr] & ($L\!\propto\!\meanSFR$) & (line shape) & [\%]\\
\midrule
$\Halpha$                       & 7   & 6.8 & 5.7 & $-17$ \\
$[$O\,\textsc{iii}$]_{5007}$    & 7   & 6.8 & 4.8 & $-30$ \\
$\CII$                          & 50  & 3.6 & 4.3 & $+17$ \\
CO(1--0)                        & 80  & 2.9 & 1.9 & $-35$ \\
\bottomrule
\end{tabular}
\caption{Mass-integrated boosts at $z=6$ under the body's
$\Lbar_\lambda\!\propto\!\meanSFR$ ansatz versus four phenomenological
$\Lbar_\lambda(\Mh)$ shapes (Yang+24 for $\Halpha$ and [O\,\textsc{iii}], Lagache+18 for $\CII$, Yang+21 for CO), all evaluated at the M26 best-fit
$(\sigPS, \tauPS)$.  The line-specific shifts are dominated by where the
$\Lbar_\lambda^2 (dn/d\log_{10}\Mh)$ weighting kernel sits relative to the
body's pivot mass: shapes that up-weight low-mass halos (Lagache+18 $\CII$)
raise $\mathcal{B}_\lambda$ via the larger $\sigPS(\Mh\!<\!10^{11}\Msun)$
in the M26 slope, while shapes that up-weight high-mass halos (Yang+21 CO,
Yang+24 short-lifetime lines) lower it.  The full spread is roughly $-35\%$
to $+17\%$ relative to the baseline values copied from Tab.~\ref{tab:boost}.}
\label{tab:appendix_boost}
\end{table}

The shifts in Table~\ref{tab:appendix_boost} are larger than the
SPS-window approximation error ($\sim\!7\%$, body Table~I caption) but
much smaller than the boost factors themselves.  We therefore consider
the body's headline numbers ($\mathcal{B}\!\sim\!7$ for short-window
and $\mathcal{B}\!\sim\!3$ for long-window tracers) robust at the
factor-of-two level against the choice of line model.

\begin{figure}[h]
\centering
\includegraphics[width=0.96\columnwidth]{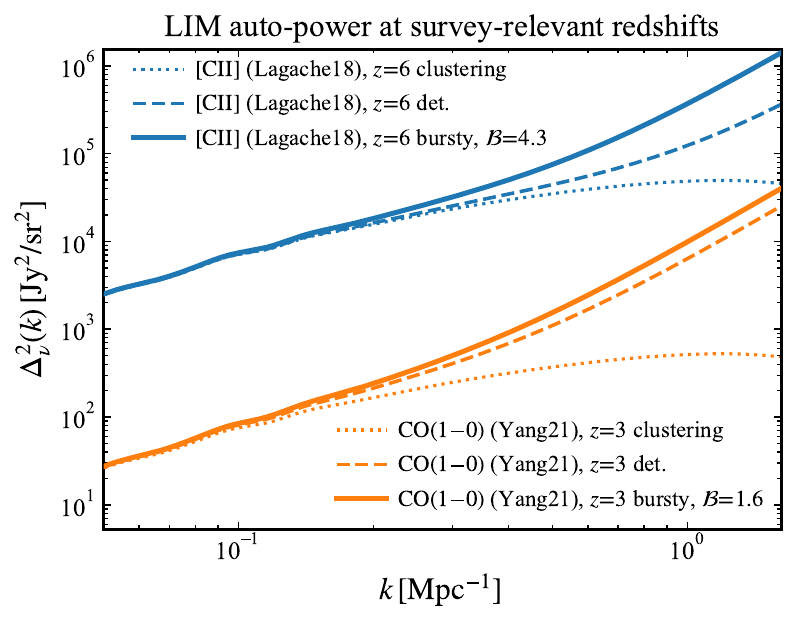}
\caption{\textbf{LIM auto-power spectra at survey-relevant redshifts.}
Solid: full $\Delta^2_\nu(k)$ with M26 burstiness ($\sigPS\!=\!2.0$,
$\tauPS\!=\!25$~Myr); dashed: deterministic prediction (no burstiness);
dotted: clustering component.  $\CII$ at $z\!=\!6$ uses the Lagache+18
prescription ($\bar I_{\CII}\!\simeq\!510$~Jy/sr,
$b_{\rm eff}\!\simeq\!2.6$, $\mathcal{B}_{\CII}\!\simeq\!4.3$); CO(1--0)
at $z\!=\!3$ uses the Yang+21 metallicity-dependent prescription
($\bar I_{\rm CO}\!\simeq\!17$~Jy/sr, $b_{\rm eff}\!\simeq\!2.7$,
$\mathcal{B}_{\rm CO}\!\simeq\!1.6$).  Mean intensities and shot-noise
normalizations are taken directly from \texttt{oLIMpus}~\cite{Libanore:2025}
under each line model.  For both lines, shot noise dominates over
clustering across most of the LIM-accessible $k$ range
($k\!\gtrsim\!0.2$~Mpc$^{-1}$ for $\CII$; $k\!\gtrsim\!0.1$~Mpc$^{-1}$
for CO), so the burstiness enhancement directly amplifies the dominant
contribution to the auto-spectrum.  The visual impact differs sharply
between the two lines: for $\CII$, the bursty curve sits a factor of
$\sim\!4$ above the deterministic prediction at high $k$, while for CO
the $\mathcal{B}_{\rm CO}\!\simeq\!1.6$ is more modest.}
\label{fig:appendix_pk}
\end{figure}

Fig.~\ref{fig:appendix_pk} extends the body's forecast to two more major LIM target lines at their experimentally relevant redshifts: $\CII$ at $z\!=\!6$ (TIME~\cite{Crites:2014}, EXCLAIM~\cite{Cataldo:2021}, CCAT/FYST~\cite{CCAT:2022}) and CO(1--0) at $z\!=\!3$ (COMAP Pathfinder~\cite{COMAP:2022}, currently observing in the COMAP-allocated band $z\!=\!2.4$--$3.4$).  $\CII$ uses the Lagache+18 prescription with boost $\mathcal{B}_{\CII}\!\simeq\!4.3$; the high mean intensity ($\bar I\!\propto\!\rho_L/\nu_{\rm rest}$~\cite{VisbalLoeb:2010} favors low-frequency lines) means shot noise dominates over clustering already at $k\!\gtrsim\!0.2$~Mpc$^{-1}$, with the bursty enhancement directly amplifying the dominant contribution to the auto-spectrum.  CO(1--0) at $z\!=\!3$ uses the Yang+21 metallicity-dependent prescription with boost $\mathcal{B}_{\rm CO}\!\simeq\!1.6$; the lower redshift puts the kernel-peak halo mass at $\sim\!10^{12}\Msun$, where $\sigPS(\Mh)$ is much smaller per the M26 mass slope, and the long Yang+21 SPS window further suppresses the boost.  The corresponding brightness temperature $\bar T_{\rm CO}\!\simeq\!0.7\,\mu$K places our prediction on the conservative end of the literature spread (e.g., the Li+16~\cite{Li:2016} fiducial gives several times brighter), with the implied bursty shot-noise power $P_{\rm shot}\!\simeq\!3\!\times\!10^2\,\mu{\rm K}^2\,{\rm Mpc}^3$ well below the COMAP Season~2 95\% upper limit~\cite{Stutzer:2024}.

\smallskip
\noindent\textit{Scatter conventions and the SFE/burstiness partition.}~%
A subtler issue arises when our boost is applied on top of line models
calibrated against the UVLF.  Codes that derive their mean SFE by
demanding the predicted UVLF match observations under an assumed scatter
$\sigma_L$ at fixed $\Mh$ (typically $\sigma_L\!=\!0$ or $\sim\!0.3$~dex)
implicitly split the observed UVLF into two contributions: the assumed
scatter and the inferred mean SFE together reproduce the data.  The M26
inference, by contrast, simultaneously fits the UVLF, galaxy clustering,
and the $\Halpha$/UV ratio with a SFR scatter $\sigPS\!\simeq\!2.0$.
The two procedures correspond to physically different ways of allocating
the same observed luminosity functions between mean-SFE and burstiness
contributions: under the $\sigma_L\!\approx\!0$ convention, the mean SFE
is large and there is no burst component, while in the M26 picture, the
mean SFE is smaller and a substantial fraction of the observed UV
emission comes from the bursty wing of the SFR distribution.  Both
conventions reproduce the UVLF; they differ in their predictions for the
LIM clustering-to-shot ratio at fixed mean intensity.

For applying our boost cleanly, this means: forecasts that calibrate
$L_\lambda(\Mh)$ phenomenologically against observed
\emph{line} luminosity functions (most COMAP, SPHEREx, EXCLAIM,
and CCAT-Prime forecasts) are not committed to a specific scatter
convention, and multiplying their predicted $\Pshot^\lambda$ by our
$\mathcal{B}_\lambda$ provides an honest estimate of the bursty
shot-noise enhancement under the M26 burstiness inference.  Forecasts
that compute $L_\lambda$ via SFR-anchored conversions on top of a
\emph{UVLF}-calibrated SFE with $\sigma_L\!=\!0$ implicitly assume a
$\sigPS\!=\!0$ universe.  Applying our boost on top of such a
prediction is internally inconsistent: it would correspond to a universe
that simultaneously satisfies $\sigPS\!\simeq\!2.0$ and the
$\sigma_L\!=\!0$ UVLF anchor, and would over-predict the LIM signal.  A
clean re-anchoring would require re-fitting the underlying SFE at the M26
mean, which is beyond the scope of this work.  Future LIM measurements
can resolve the partition empirically: the clustering-to-shot ratio,
which is independent of the absolute amplitude calibration, is set by
$\mathcal{B}_\lambda$ for a given $\Lbar_\lambda(\Mh)$ shape and is the
natural observable for distinguishing the two
conventions~\cite{YueFerrara:2019}.

We also note that we have not imposed an upper cap on $\sigPS(\Mh)$ at low
halo mass.  Sufficiently large bursts ($\sigPS\!\gtrsim\!3$) are physically
self-limiting: such amplitudes would consume a low-mass halo's available
cold gas on the burst-coherence timescale, requiring gas depletion and
SN/radiative feedback to truncate the burst before this regime is reached.
The lognormal description of the SFR PDF is also expected to break down
here. At $z\!=\!6$ the kernel-integrated boost is dominated by halos with
$\sigPS\!\lesssim\!2.5$, where neither concern applies. At higher redshift,
lines with sublinear or extended low-mass $\Lbar_\lambda(\Mh)$ shapes
(notably $\CII$ under Lagache+18) develop low-mass kernel tails reaching
into the $\sigPS\!\gtrsim\!3$ regime. Imposing a hard cap $\sigPS\!\leq\!3$
shifts $\mathcal{B}_{\CII}$ (Lagache+18) from $-0.9\%$ at $z\!=\!6$ to $-5.1\%$
at $z\!=\!10$, while $\Halpha$, $[\textrm{O\,\textsc{iii}}]$, and CO are
essentially unaffected (cap shifts $\lesssim\!0.5\%$ across all redshifts).
The redshift-tomography prediction in Fig.~\ref{fig:redshift}, which uses
the body's $\Lbar\!\propto\!\meanSFR$ assumption, is robust against this
systematic at the sub-percent level for $\Halpha$ across the full
$z\!=\!5$--$15$ range.

\section*{Appendix B: Derivation of $V_\lambda(\Mh)$ and verification}

Here we derive the closed-form expression for the per-halo luminosity variance Eq.~\eqref{eq:Vexact} from the M26 SFR model, verify it against Monte Carlo realizations of the OU process, and show how the result compares with the trivial $e^{\sigma_x^2}$ stochasticity prescription used in most LIM literature.

\smallskip
\noindent\textit{Why the Green function survives only in the second moment.}
The mean line luminosity is $\Lbar_\lambda(\Mh, t_{\rm obs})\!=\!\int dt_a\,G_\lambda(t_a)\,\langle\dot M_\star\rangle(\Mh, t_{\rm obs}\!-\!t_a)$, where the lognormal expectation $\langle\dot M_\star\rangle\!=\!\meanSFR\,\langle e^{x-\sigma_x^2/2}\rangle$ trivially equals the UVLF-anchored value, since $\langle e^x\rangle\!=\!e^{\sigma_x^2/2}$ exactly cancels the $e^{-\sigma_x^2/2}$ offset in the SFR definition Eq.~\eqref{eq:SFR_lognormal}. Throughout this work we adopt the time-stationary approximation $\langle\dot M_\star\rangle(\Mh, t_{\rm obs}\!-\!t_a)\!\approx\!\langle\dot M_\star\rangle(\Mh, t_{\rm obs})$ across the support of $G_\lambda$, justified at the few-percent level at $z\!\sim\!6$ (cosmic time changes by $\sim\!10\%$ over $100$~Myr; SFRs in our halo-mass range change by less). Under this approximation the Green function in the first moment reduces to a line-dependent normalization, $\Lbar_\lambda(\Mh, t_{\rm obs})\!\approx\!\kappa_\lambda\langle\dot M_\star\rangle(\Mh, t_{\rm obs})$ with $\kappa_\lambda\!=\!\int dt_a\,G_\lambda(t_a)$. The second moment is fundamentally different because $\langle e^{x_1+x_2}\rangle\!=\!e^{\sigma_x^2+\xi_x(t_1-t_2)}$ leaves the time correlator $\xi_x$ uncancelled. The mean-anchoring choice thus routes burstiness exclusively into $\langle L_\lambda^2\rangle$; relaxing the time-stationary approximation introduces a deterministic look-back correction to $\Lbar_\lambda$ that affects the clustering and shot-noise baselines together, leaving the boost $\mathcal{B}_\lambda\!\equiv\!\Pshot/\Pshot|_{\rm det}$ unchanged.

\smallskip
\noindent\textit{From SFR statistics to $V_\lambda$.}
The line luminosity is $L_\lambda(\Mh)\!=\!\int_0^\infty dt_a G_\lambda(t_a)\dot M_\star(\Mh,t_{\rm obs}\!-\!t_a)$.  Defining the window $W_\lambda(\Mh,t_a)\!\equiv\!G_\lambda(t_a)\meanSFR(\Mh,t_{\rm obs}\!-\!t_a)$ and writing $L_\lambda\!=\!\int W_\lambda y\,dt_a$ with $y\!\equiv\!e^{x-\sigma_x^2/2}$ (so $\langle y\rangle\!=\!1$), the luminosity correlator is set by the lognormal-mapped Gaussian correlator~\cite{Xavier:2016,Caplar:2019}
\be
\xi_y(\Delta t) = \exp[\xi_x(\Delta t)] - 1,
\label{eq:xi_y_app}
\ee
which is \emph{not} well approximated by $\xi_x$ when $\sigma_x^2\!\sim\!\mathcal{O}(1)$.  The dimensionless luminosity variance at fixed $\Mh$ is then
\be
V_\lambda(\Mh) = \frac{\int dt_1 dt_2\, W_\lambda(t_1) W_\lambda(t_2)\,\xi_y(t_1\!-\!t_2)}{[\int dt_a\,W_\lambda]^2}.
\label{eq:V_def_app}
\ee
$V_\lambda$ depends on $W_\lambda$ only through its autocorrelation:
\be
V_\lambda = \int_0^\infty\!\!du\,\Bigl[2\!\int_0^\infty\!\!W_\lambda(t)W_\lambda(t\!+\!u)\,dt\Bigr][e^{\xi_x(u)}\!-\!1].
\label{eq:V_kernel_autocorr}
\ee
A top-hat of width $t_\lambda$ has the same autocorrelation peak and integrated coherence as an exponential of e-folding $t_e\!=\!t_\lambda/2$, which is the matching prescription we adopt for our $t_\lambda$ values.  For $\Halpha$ and UV, where the Green's function is set by stellar physics, direct numerical comparison of the top-hat and exponential forms via Eq.~\eqref{eq:V_def_app} confirms agreement at the $\lesssim\!5\%$ level in $\sigxiline^2(\Mh)\!=\!\ln[1+V_\lambda(\Mh)]$ across the relevant mass range.

For an OU correlator $\xi_x(\Delta t)\!=\!\sigma_x^2 e^{-|\Delta t|/\tauPS}$, the spectral density of $y$ admits a closed form. Expanding $\xi_y$ as a Taylor series and Fourier-transforming term by term gives
\be
P_y(\omega) = \sum_{n=1}^{\infty}\frac{(\sigma_x^2)^n}{n!}\,\frac{2n\tauPS}{n^2+(\omega\tauPS)^2},
\label{eq:Py_OU_app}
\ee
each term being a Lorentzian of width $n/\tauPS$.  The series converges very rapidly because the $n$-th term scales as $(\sigma_x^2)^n/n!$, so $P_y$ is an entire function of $\sigma_x^2$ with no convergence concern at any amplitude we encounter; the $n\!\geq\!2$ terms are the non-Gaussian enhancement, and dominate $V_\lambda$ when $\sigma_x^2\!\sim\!\mathcal{O}(1)$.  Substituting Eq.~\eqref{eq:Py_OU_app} into a top-hat window of width $t_\lambda$ and performing the Fourier transform yields the body's Eq.~\eqref{eq:Vexact}, which is exact for any $\sigma_x^2$.

For $\tauPS\!\gg\!t_\lambda$ the integrand approaches $e^{\sigma_x^2}\!-\!1$ uniformly, giving $V_\lambda\!\to\!e^{\sigma_x^2}\!-\!1$.  For $\tauPS\!\ll\!t_\lambda$ the integral is dominated by $s\!\lesssim\!\tauPS$, yielding $V_\lambda\!\simeq\!(2\tauPS/t_\lambda)\!\int_0^1\!du\,(e^{\sigma_x^2 u}\!-\!1)/u$ via the substitution $u\!=\!e^{-s/\tauPS}$ in the leading-order asymptotic expansion, which is parametrically much larger than the $(\tauPS/t_\lambda)\sigma_x^2$ scaling that would obtain if one substituted $\xi_x$ for $\xi_y$.  As a quantitative check, for UV continuum at $\Mh\!=\!10^{11}\Msun$ (where $\sigma_x^2\!=\!2.0$ at our central values), Eq.~\eqref{eq:Vexact} gives $V_\lambda\!\simeq\!1.5$ versus a linearized $V_\lambda\!\simeq\!0.76$; at $\Mh\!=\!10^{10}\Msun$ ($\sigma_x^2\!\simeq\!3.1$) the exact $V_\lambda\!\simeq\!3.9$ exceeds the linearized estimate $V_\lambda\!\simeq\!1.2$ by more than a factor of three.  This non-Gaussian enhancement at large $\sigma_x^2$ is what drives the boost values reported in the body.

\smallskip
\noindent\textit{Monte Carlo verification.}
We verified Eq.~\eqref{eq:Vexact} against Monte Carlo realizations of the OU process for representative cases ($t_\lambda\!=\!7,\,50,\,100$~Myr at $\sigPS\!=\!2.0$, $\tauPS\!=\!25$~Myr).  Across 10 batches of $10^6$ realizations each, the closed form lies within the sampling spread of the MC estimator at all three timescales (no systematic bias detected); the per-batch scatter of the MC estimator itself is large ($\sim\!1$--$3\%$, growing with $t_\lambda$), reflecting the heavy-tailed sampling of the lognormal $L_\lambda$ rather than any deficiency of Eq.~\eqref{eq:Vexact}.

\smallskip
\noindent\textit{Comparison with the trivial $e^{\sigma_x^2}$ prescription.}
Earlier LIM forecasts often introduce per-galaxy stochasticity by simply multiplying the deterministic shot noise by $e^{(\sigma_L\ln 10)^2}$ with a phenomenological $\sigma_L\!\sim\!0.3$~dex calibrated against simulations, without convolving against the SPS Green's function.  Our framework departs from this in three ways: (i) the M26-inferred $\sigPS$ is much larger than $0.3$~dex; (ii) the mass dependence weights the integral toward lower-mass halos where $\sigPS$ is even larger; and (iii) the Green-function convolution that defines $V_\lambda$ encodes the time response of each line.  Effects (i) and (ii) push the boost \emph{up} relative to the literature prescription; effect (iii) pushes it \emph{down} relative to a naive $e^{\sigma_x^2}\!-\!1$ result using the M26 amplitude, because time-averaging across a broad SPS window partially smooths the bursts.  For long-window tracers (UV, CO, [\textsc{C\,ii}]) the latter effect can be substantial, reducing the per-halo boost by factors of a few relative to the $\tauPS\!\gg\!t_\lambda$ asymptote.  The two corrections combine to give the modest factor $\mathcal{B}_\lambda\!\sim\!1.5$--$3$ enhancement of long-window boosts relative to the $\sigma_L\!=\!0.3$~dex literature reference (compared with $\sim\!4$--$5$ for $\Halpha$, where the Green-function effect is small); the dotted reference line in Fig.~\ref{fig:robustness} marks the literature value.

\end{document}